\documentclass{appolb}

\usepackage{graphicx}
\usepackage{subfigure}
\usepackage{bm}

\makeatletter
\newcommand\erfc{\mathop{\operator@font erfc}\nolimits}
\def\slashchar#1{\setbox0=\hbox{$#1$}
   \dimen0=\wd0 \setbox1=\hbox{/} \dimen1=\wd1
   \ifdim\dimen0>\dimen1 \rlap{\hbox to \dimen0{\hfil/\hfil}} #1
   \else  \rlap{\hbox to \dimen1{\hfil$#1$\hfil}} / \fi}

\makeatother

\begin{document}
\bibliographystyle{h-elsevier3} 
\title{Viscosity and boost invariance at RHIC and LHC\thanks{Talk presented at
the Cracow Epiphany Conference on LHC Physics,
4 - 6 January 2008, Cracow, Poland. Supported by 
Polish Ministry of Science and Higher Education under
grant N202~034~32/0918.}}
\author{Piotr Bo\.zek
\footnote{email:~Piotr.Bozek@ifj.edu.pl}
 \address{Institute of Nuclear Physics PAN,
PL-31342 Krak\'ow, Poland}
\and
\address{Institute of Physics, Rzesz\'ow University, 
PL-35959 Rzesz\'ow, Poland} }
%

\maketitle

\begin{abstract}
We consider the longitudinal hydrodynamic evolution of 
the fireball created in a relativistic heavy-ion collision.  Nonzero 
shear viscosity reduces the colling rate of the system and hinders the 
acceleration of the longitudinal flow. As a consequence, the 
initial energy density needed to reproduce the experimental data at RHIC 
energies is significantly 
reduced. At LHC energies, we expect that shear viscosity 
helps to conserve a Bjorken plateau in the rapidity distributions during the 
expansion.
\end{abstract}



\section{Introduction\label{sec:intro}}

In heavy-ion collisions at relativistic energies a fireball 
of dense and hot matter is created. In later stages, 
 the fireball expands and eventually hadronizes. In order to
deduce the properties of that medium, including a possible phase
 transition to the quark-gluon plasma, a careful modelling of the 
evolution of the system is needed. With the assumption of 
local thermal equilibrium, the description of the dynamics of
 the system can be undertaken within the
relativistic hydrodynamics 
 \cite{Kolb:2003dz}. Hydrodynamic calculations reproduce the spectra of
 particles in the transverse momentum, the collective flow, and 
 the Hanbury-Brown-Twiss radii measured at Relativistic Heavy Ion Collider
(RHIC) \cite{Chojnacki:2007rq}.
 Such calculations are  essential in determining the correct 
equation of state of  the matter under such extreme conditions, the initial 
conditions of the system  and the freeze-out time. 

Until recently,  hydrodynamic calculations were performed assuming  
the  applicability of the ideal fluid limit. It means that local equilibration
 processes are instantaneous, and no entropy is produced in the 
hydrodynamic stage. Relaxing the ideal fluid assumption amounts 
to the introduction of viscous effects in the  evolution. 
For the fireball created at RHIC and Large Hadron Collider (LHC) energies,
 the most important dissipative effect in the hydrodynamic
evolution is due to the shear viscosity
\cite{Song:2007fn,Teaney:2003kp,Baier:2006sr,Baier:2006gy,Chaudhuri:2006jd,Muronga:2004sf}. A nonzero shear viscosity coefficient causes 
the saturation of the elliptic flow, a stronger transverse flow and can lead 
to a significant entropy production. Most of the 
calculations of the hydrodynamic evolution of the fireball with viscosity are 
done in the transverse directions only, with boost-invariance assumed in the beam (longitudinal) direction. However, 
shear viscosity is also important for the 
longitudinal expansion of the fireball \cite{Bozek:2007qt}. 
In the following,
 we present results
 of calculations in a 1+1 dimensional geometry of a  non-boost-invariant expanding 
fluid
 with viscosity. The cooling rate and the acceleration of the
 longitudinal flow are reduced, and the 
entropy is produced in the expansion.

\section{Longitudinal expansion with viscosity}
Relativistic hydrodynamics with viscosity can be formulated consistently, without violating the causality \cite{IS}. The hydrodynamic equations
\begin{equation}
\partial_\mu T^{\mu\nu}=0
\end{equation}
are modified; the energy-momentum tensor $T^{\mu\nu}=T^{\mu\nu}_{ideal}
+\pi^{\mu\nu}$ is composed of the energy-momentum tensor of an ideal fluid and 
a stress tensor $\pi_{\mu\nu}$ describing  deviations from local equilibrium.
Assuming the dependence of the densities only on the time $t$ 
and the longitudinal coordinate $z$ and restricting the flow velocity 
$u^\mu=(\gamma,0,0,\gamma v)$ only to the beam direction, one can write  
the hydrodynamic equations for a fluid with shear
 viscosity as \cite{Bozek:2007qt}
\begin{eqnarray}
(\epsilon+p)DY&=& -{\cal K}p+\Pi DY + {\cal K} \Pi \nonumber \\
D\epsilon&=& (\epsilon +p) {\cal K} Y -\Pi {\cal K} Y\nonumber \\
D\Pi&=& (\frac{4}{3}\eta {\cal K}Y-\Pi)/\tau_\pi  \ ,
\label{eq:eqsolv}
\end{eqnarray}
where $D=u^\mu\partial_\mu=\cosh(Y-\theta)\partial_\tau
+\frac{\sinh(Y-\theta)}{\tau}\partial_\theta$, 
${\cal K}=\sinh(Y-\theta)\partial_\tau+\frac{\cosh(Y-\theta)}
{\tau}\partial_\theta $. $\theta=\frac{1}{2}\ln\left(\frac{t+z}{t-z}\right)$ 
is the space-time rapidity and
 $Y=\frac{1}{2}\ln\left(\frac{E+zp_z}{E-p_z}\right)$ is the 
kinematic rapidity of a fluid element. Eqs. (\ref{eq:eqsolv}) involve 
four unknown function of $\theta$ and of the proper time $\tau=\sqrt{t^2-z^2}$~:
 the energy density $\epsilon$, the pressure $p$, the rapidity
 $Y$ of the fluid element,  and the shear  correction 
$\Pi$ (in the 1+1 dimensional  geometry the stress tensor $\pi^{\mu\nu}$ 
reduces to one independent scalar function $\Pi$). One additional
 relation is given by the equation of state connecting $\epsilon$ and $p$.
We use a realistic equation of state by Chojnacki and 
Florkowski combining   lattice QCD results and 
a hadron gas model  \cite{Chojnacki:2007jc}. The ratio 
of the shear viscosity coefficient $\eta$ to the entropy of the fluid $s$ 
 is not known. We perform calculation for several values $\eta/s=0.1$-$0.3$ 
of this parameter. For the relaxation time of dissipative corrections we take 
$\tau_\pi=6\eta/Ts$ \cite{Venugopalan:1992hy}. The hydrodynamic equations
(\ref{eq:eqsolv})
 are solved in the $\tau-\theta$ plane
starting from initial conditions at $\tau_0=1$fm/c
\begin{eqnarray}
\epsilon(\tau_0,\theta)&=&\epsilon_0 \exp\left(-\theta^2/(2\sigma^2)\right)
\nonumber \\
Y(\tau_0,\theta)&=&\theta \ ,
\end{eqnarray}
i.e. a Gaussian initial energy density profile and the Bjorken
 collective longitudinal flow of the fluid.

\section{RHIC results}

Eqs. (\ref{eq:eqsolv}) are solved numerically and the freeze-out 
hypersurface of constant temperature $T_f=165$MeV is extracted.
\begin{figure}
\includegraphics[width=.7\textwidth]{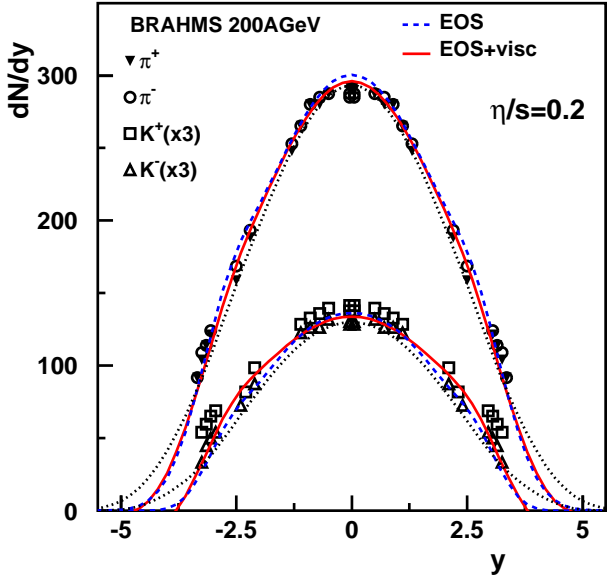}
\caption{ Rapidity distributions for pions and kaons calculated 
with the ideal fluid hydrodynamics (dashed line) and using the 
viscous hydrodynamics with $\eta/s = 0.2$ (solid line) \cite{Bozek:2007qt}.
 The dotted line denotes the results of the viscous hydrodynamic evolution,
 but neglecting the viscous corrections to the particle emission at 
freeze-out (Eq. \ref{eq:dndydissi}). Data are from the BRAHMS Collaboration 
\cite{Bearden:2004yx}.}
\label{fig:dndy}
\end{figure}
At the freeze-out hypersurface, hadrons are emitted 
according to the Cooper-Frye formula \cite{Cooper:1974mv,Teaney:2003kp}
\begin{equation}
\frac{dN_{visc}}{dy}=\frac{dN}{dy}+\frac{d\delta N}{dy} \ \ .
\label{eq:dndy}
\end{equation}
with
the usual thermal emission contribution of the form
\begin{eqnarray}
\frac{dN}{dy}&=& \frac{S}{4\pi^2}\int_{-\theta_{max}}^{\theta_{max}} 
\left( \tau(\theta) \cosh(y-\theta)-\tau^{'}(\theta)\sinh(y-\theta)\right)
\nonumber \\ & & 
(2 m \xi +2 \xi^2 +m^2)\xi \nonumber \\ & & 
\exp\left( -\frac{m\cosh(y-Y_f(\theta))}{T_f}  \right) d\theta \ , 
\label{eq:dndyfree}
\end{eqnarray}
\begin{figure}
\includegraphics[width=.7\textwidth]{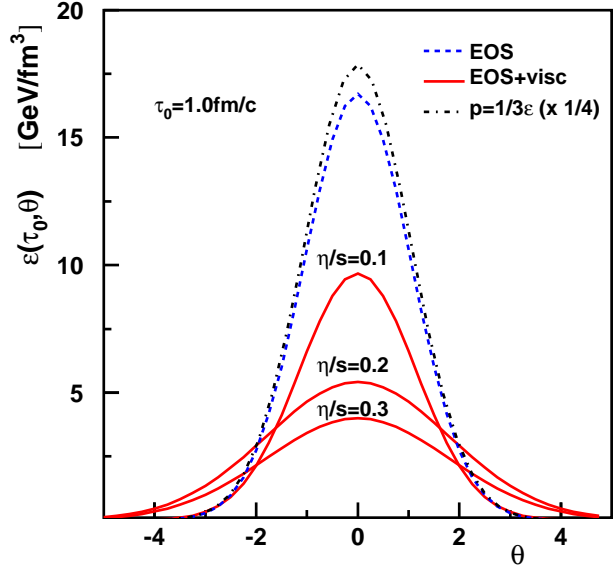}
\caption{ Initial energy density distributions for 
the ideal fluid hydrodynamic evolution with a realistic equation of
 state (dashed line), for  several viscous hydrodynamic evolutions 
(solid lines), and for the ideal fluid with 
a relativistic gas equation of state 
(dashed-dotted line) \cite{Bozek:2007qt}.
 All distributions after the hydrodynamic evolution 
and freeze-out give  pions distributions close to the BRAHMS measurements 
\cite{Bearden:2004yx}. }
\label{fig:eini}
\end{figure}
where $m$ is the meson mass, $Y_f(\theta)=Y(\tau(\theta),\theta)$ is 
the fluid rapidity at the freeze-out hypersurface, and
\begin{equation}
\xi=\frac{T_f}{\cosh(y-Y_f(\theta))} \ ;
\end{equation}
and an additional term due to the viscous corrections \cite{Teaney:2003kp}
\begin{eqnarray}
\frac{d\delta N}{dy}& &= \frac{S}{4\pi^2}\int_{-\theta_{max}}^{\theta_{max}} 
\left( \tau(\theta) \cosh(y-\theta)-\tau^{'}(\theta)\sinh(y-\theta)\right)
\nonumber \\ & & 
\left[12 \xi^5 +5\xi^3m^2+12 \xi^4 m + \xi^2 m^3 -\sinh(y-Y_f(\theta)) 
\right. \nonumber \\ & &  \left.
(24 \xi^5 +12 \xi^3 m^2 +24\xi^4 m +4 \xi^2 m^3 +\xi m^4) \right]
\nonumber \\ & &
\frac{\Pi}{2T^2 (\epsilon+p)}
\exp\left( -\frac{m\cosh(y-Y_f(\theta))}{T_f}  \right) d\theta \ .
\label{eq:dndydissi}
\end{eqnarray}

At the freeze-out temperature, most of the pions and 
kaons come from secondary decays of heavier resonances. 
This effect is taken into account by a factor, equal to
 the ratio of all mesons to primary mesons\cite{Torrieri:2004zz}, 
 multiplying the calculated distributions. 
The parameters of the initial distribution $\sigma$ and $\epsilon_0$ are 
adjusted to reproduce the pion and kaon rapidity distributions measured by the 
BRAHMS Collaboration \cite{Bearden:2004yx}. 
\begin{figure}
\includegraphics[width=.7\textwidth]{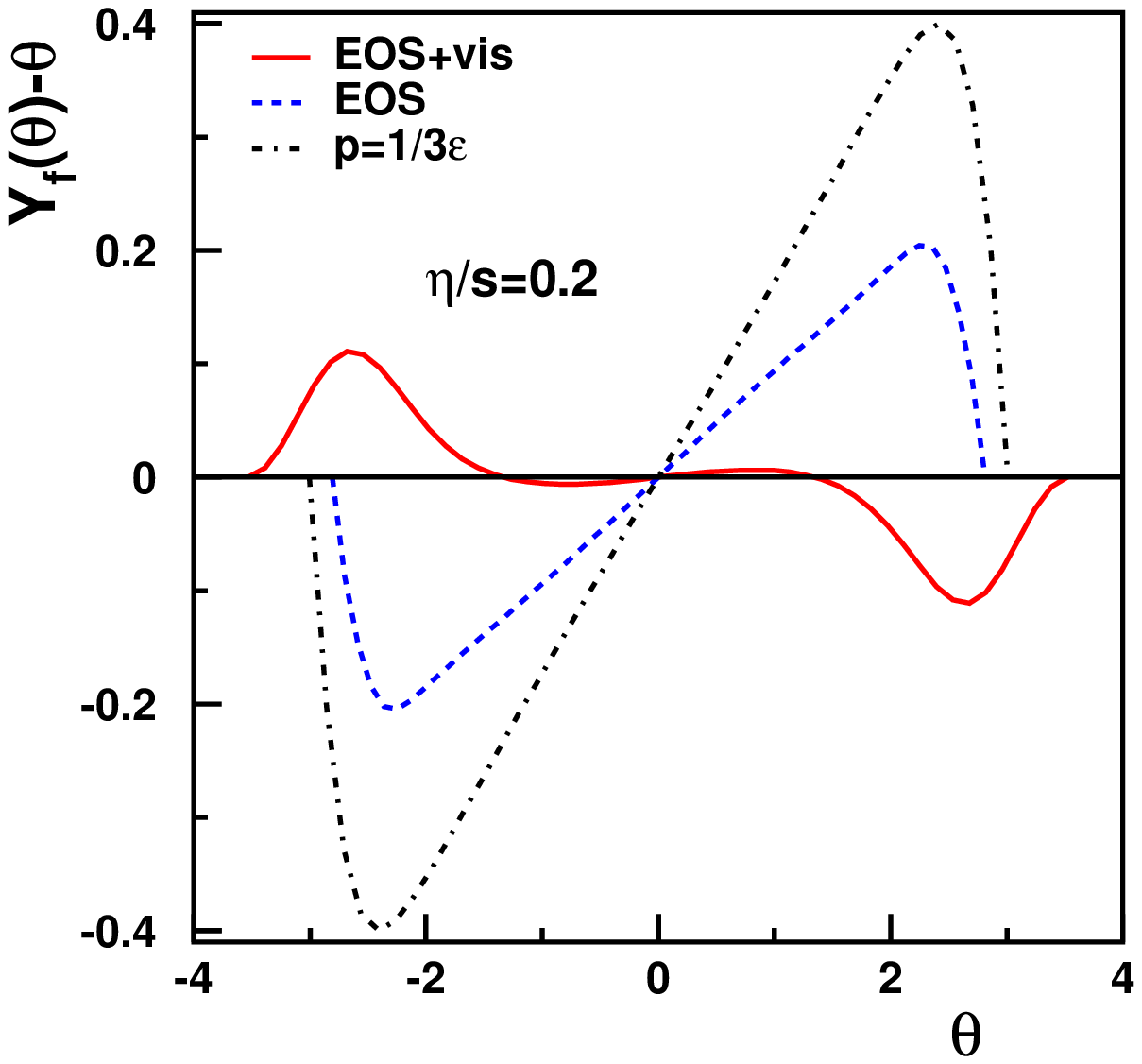}
\caption{Difference between the  flow rapidity 
of the fluid and the Bjorken flow  taken at the freeze-out hypersurface, 
calculated for the evolution with the
shear viscosity coefficient $\eta/s = 0.2$ (solid line), for the ideal
 fluid with a  realistic equation of state (dashed line) and 
for the ideal fluid with a
 relativistic gas equation of state  (dashed-dotted line) \cite{Bozek:2007qt}.}
\label{fig:flowfreeze}
\end{figure}
With the increase of the  shear viscosity of the fluid, two effects can be observed (Fig. \ref{fig:eini}) in the retuned initial state of the fireball~:
\begin{itemize}
\item a reduced initial energy density of the fireball
\item and a small increase of the width of the initial distribution $\sigma$ 
with increasing viscosity.
\end{itemize}

The two above mentioned effects are related to the change of the 
dynamics induced by the 
shear viscosity. The reduction of the initial energy density at 
 central rapidities is related to the reduced cooling rate in the 
viscous evolution.
It originates 
 from the smaller work of the fluid in the longitudinal viscous
expansion, due to the change  of the effective pressure from $p$ to $p-\Pi$ 
(first equation in (\ref{eq:eqsolv})). The second effect is related 
to the acceleration of the
 longitudinal flow, as given by the second equation in (\ref{eq:eqsolv}). 
Gradients 
of the pressure in the longitudinal direction cause the acceleration of the 
flow, which becomes faster than the Bjorken one \cite{Satarov:2006iw}.
In the viscous evolution, the  gradients of the pressure $p$
 are reduced by the gradients of the shear correction $\Pi$.
As a result, the flow at the freeze-out is still Bjorken-like for 
the viscosity $\eta/s=0.2$ (Fig. \ref{fig:flowfreeze}),
 and is significantly slower than
 for the ideal fluid. 
Fast moving fluid elements emit hadrons far in the forward and 
background rapidities, to counteract this effect a narrower initial energy 
density distribution in rapidity must be assumed for the ideal fluid.

\section{Predictions for LHC}

\begin{figure}
\includegraphics[width=.7\textwidth]{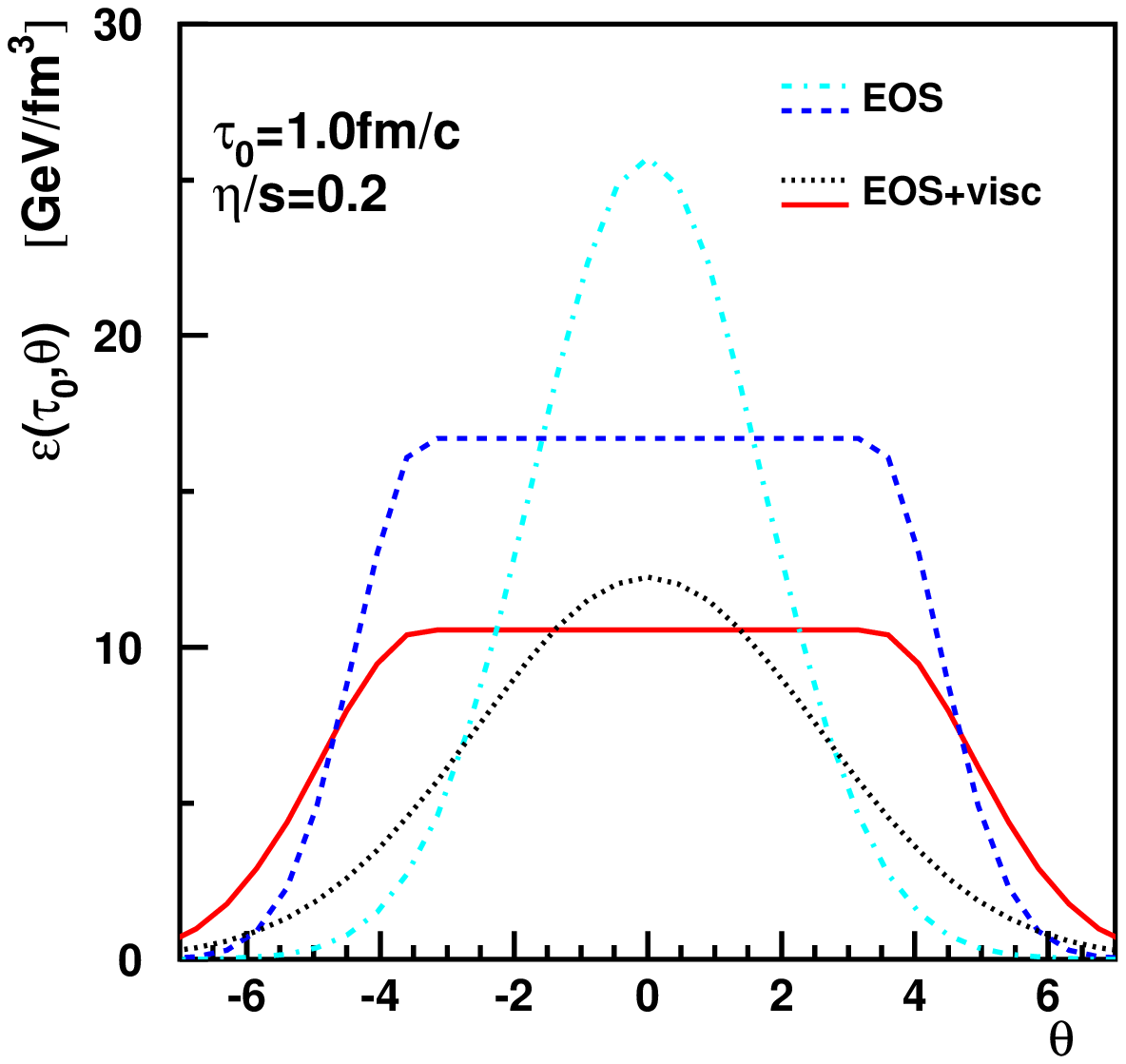}
\caption{ Initial energy density distributions for 
 ideal fluid hydrodynamic evolutions (dashed and dashed-dotted lines)
 and for  viscous hydrodynamic evolutions 
(solid and dotted lines).
 All distributions after the hydrodynamic evolution 
and freeze-out give the same pion density at central rapidity.}
\label{fig:einil}
\end{figure}

\begin{figure}
\includegraphics[width=.7\textwidth]{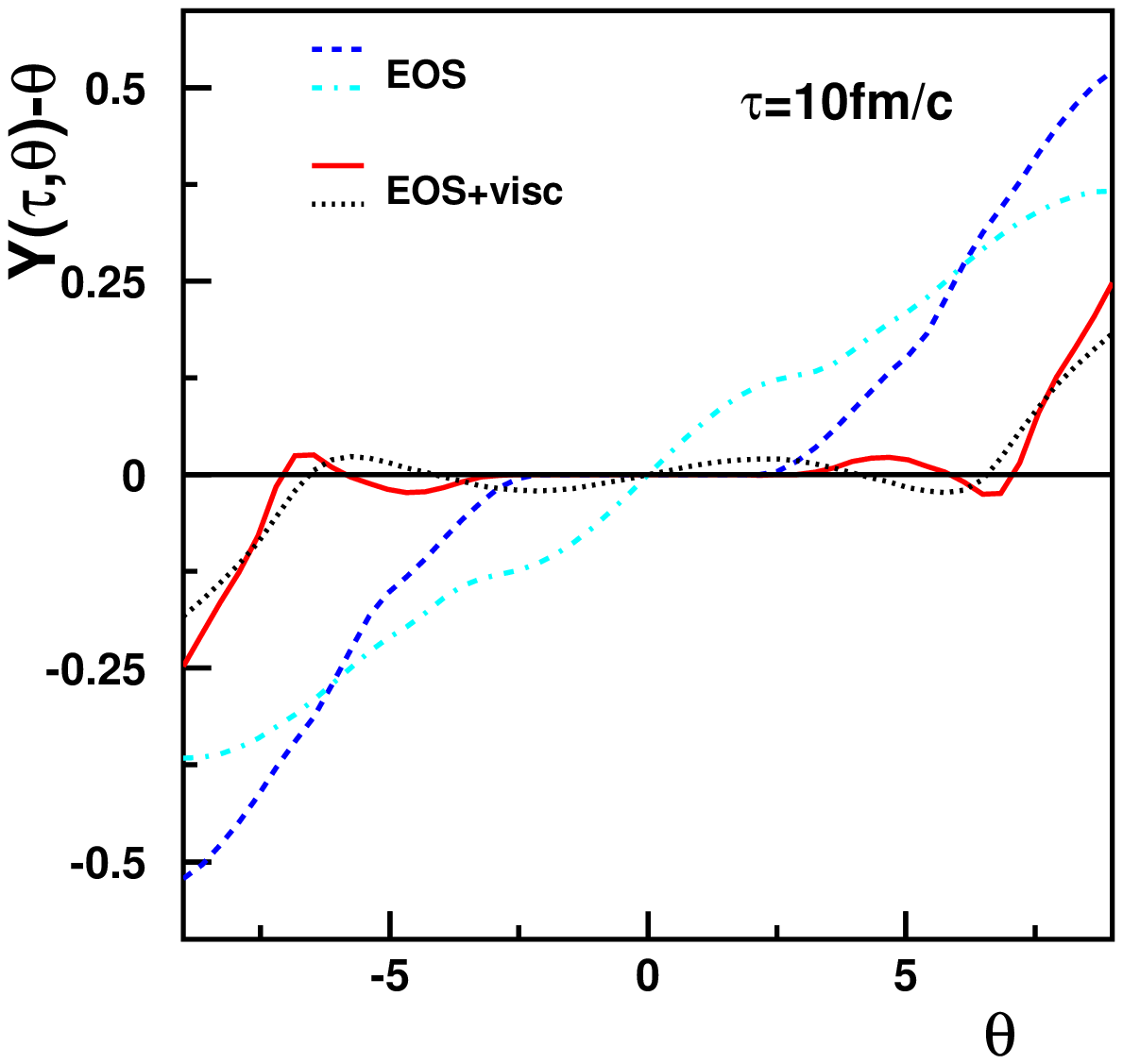}
\caption{Difference between the  flow rapidity 
of the fluid and the Bjorken flow taken at $\tau=10$fm/c,
 calculated for  evolutions with 
shear viscosity coefficient $\eta/s = 0.2$ (solid and dotted lines)
 and for ideal
 fluid  evolutions (dashed and dashed-dotted lines)
 for LHC initial conditions (Fig. \ref{fig:einil}).}
\label{fig:flowl}
\end{figure}

For the forthcoming LHC Pb-Pb heavy-ion experiments, we assume arbitrarily
that the particle multiplicities at central rapidities would increase
 twice compared to Au-Au at the highest RHIC energies. This constraint serves
to fix the initial energy density $\epsilon_0$ for the hydrodynamic evolution.
At RHIC energies, experimental results show that a Bjorken 
scaling plateau at central rapidities, if existing at all, is very narrow 
\cite{Hirano:2002ds,Satarov:2006iw,Bozek:2007qt}. 
At LHC energies, we consider two different scenarios~: a
 Gaussian initial energy density distribution in the space-time rapidity 
(similar as for RHIC), 
or a 
distribution with a plateau of width $2\sigma_p$ at central rapidities
\begin{equation}
\epsilon(\tau_0,\theta)=\epsilon_0 
\exp\left(-(\theta-\sigma_p)^2\Theta(|\theta|-
\sigma_p)/2\sigma^2\right) \ .
\label{eq:einilhc}
\end{equation}
The initial energy density distributions for the two scenarios  are
 shown in Fig. \ref{fig:einil},
both for the ideal fluid and for the viscous hydrodynamics with $\eta/s=0.2$. 
The initial energy density is smaller for 
nonzero viscosity. The reduction of the initial density  accounts for 
 the slower cooling and the entropy production 
in the later viscous evolution.  

The scenario with a Bjorken scaling plateau in the range of 
$8$ units of central rapidities leads to different results   than the one with 
the Gaussian initial density profile. Within the plateau region, the Bjorken
 scaling flow is stable, the pressure gradient in space-time rapidity is zero.
During the evolution, the plateau region is becoming narrower, 
as the gradient of the pressure at the edges of the plateau
 starts to destroy the 
scaling form of the density and of the flow. If the 
shear viscosity is large, 
the rate at which the Bjorken scaling region 
is reduced is smaller than in the ideal fluid evolution. As a consequence, 
at the freeze-out, a substantial region with Bjorken scaling of the density 
and flow survives, if it is present in the initial state and if the 
shear viscosity is large (Fig. \ref{fig:flowl}).
For the final meson distribution  a
 plateau at central rapidities is possible only
 for the viscous evolution from an initial condition
 with a plateau in the initial energy density distribution
 (solid line in Fig. \ref{fig:einil}).

\section{Summary}

We calculate the evolution of the matter created in relativistic 
heavy-ion collisions in the longitudinal direction 
both for the ideal and for the viscous fluid hydrodynamics \cite{Bozek:2007qt}.
Starting with Gaussian profiles of the energy density in 
space-time rapidity and with a Bjorken scaling longitudinal flow, 
the hydrodynamic evolution  reduces the energy density and
 accelerates the longitudinal flow. The rate of these processes is governed by 
gradients of the effective pressure $p-\Pi$, therefore nonzero
shear viscosity corrections $\Pi$
 reduce the cooling rate and, at the same time, make
 the longitudinal flow to stay closer to the initial Bjorken scaling flow. At RHIC energies, comparison to meson rapidity distributions
 of the BRAHMS Collaboration \cite{Bearden:2004yx} allows to constraint 
the parameters of the initial state. 
Depending on the value of the shear viscosity 
coefficient one gets a reduction of the initial energy 
density by a factor $2$-$3$ for $\eta/s=0.1$-$0.2$. At LHC energies, the same
effects take place for similar Gaussian initial conditions. 
Assuming an initial distribution with a plateau at central rapidities, 
where the Bjorken scaling solution applies, the dynamics is different. Within 
the plateau region the Bjorken flow is stable, 
both in the ideal and viscous hydrodynamics. Nonzero viscosity helps to preserve the Bjorken plateau in a wider region of rapidities trough the evolution; 
reduced
gradients in the hydrodynamic equations make smaller the rate 
at which the Bjorken plateau diminishes. For $\eta/s=0.2$, the  plateau region 
remaining till freeze-out is wide enough to survive the hadron emission
 process, and could be visible as a plateau in the final meson distribution in
 the kinematic rapidity.

\bibliography{hydr}

\end{document}